\documentclass[11pt,a4paper]{article}
\usepackage{jheppub}

\usepackage[utf8]{inputenc}
\usepackage{amssymb}
\usepackage{amsmath}
\usepackage{amsfonts}
\usepackage{graphicx}
\usepackage{color}
\usepackage{xspace}
\usepackage{hyperref}
\usepackage[normalem]{ulem}

\usepackage[section]{placeins}
\usepackage{afterpage}

\usepackage{float}
\usepackage{slashed}
\usepackage{ulem}
\usepackage{appendix}

\usepackage{cancel}

\usepackage{multirow,rotating}
\usepackage[dvipsnames]{xcolor}

\usepackage{orcidlink}

\title{Searching for long-lived light neutralinos and axionlike particles at the SHiNESS experiment}

\author[\,a,\,b,\,c]{Zeren Simon Wang\,\orcidlink{0000-0002-1483-6314}}
\emailAdd{wzs@mx.nthu.edu.tw}
\affiliation[a]{School of Physics, Hefei University of Technology, Hefei 230601, China}
\affiliation[b]{Department of Physics, National Tsing Hua University, Hsinchu 300, Taiwan}
\affiliation[c]{Center for Theory and Computation, National Tsing Hua University, Hsinchu 300, Taiwan}

\author[\,a, 1]{Yu Zhang\,\orcidlink{0000-0001-9415-8252}\note{Corresponding author.}}
\emailAdd{dayu@hfut.edu.cn}

\author[\,d, 2]{Wei Liu\,\orcidlink{0000-0002-3803-0446}\note{Corresponding author.}}
\emailAdd{wei.liu@njust.edu.cn}
\affiliation[d]{Department of Applied Physics and MIIT Key Laboratory of Semiconductor Microstructure and Quantum Sensing, Nanjing University of Science and Technology, Nanjing 210094, China}

\abstract{Recently Soleti et al.~[\textit{JHEP03(2024)148}] proposed a new experiment called SHiNESS at the upcoming European Spallation Source (ESS) facility, making use of the 2-GeV proton beam there impinging on a fixed target, in order to search for hidden sterile neutrinos that could lie in different mass ranges and arise with distinct signatures. Such signatures include excesses in electron-positron pairs that may originate from displaced decays of long-lived particles (LLPs). At the ESS, the dominant sources of such LLPs are decays at rest of $\pi^+$ mesons and $\mu^+$ leptons. We choose to investigate theoretical scenarios of long-lived light neutralinos in the R-parity-violating supersymmetry and long-lived weak-violating electrophilic axionlike particles, as these LLPs can be produced from decays of $\pi^+$ and $\mu^+$. Since the $\pi^+$'s {and $\mu^+$'s} decay at rest at the ESS, we compute the spectra of the therefrom produced  LLPs, and thus estimate the expected sensitivity reach of SHiNESS to the LLPs in these two scenarios. Our calculation shows that in multiple relevant benchmark scenarios, SHiNESS can probe large parameter regions of these models beyond the existing bounds, in just a couple of years of data-collection time.}

\begin{document}


	
	\maketitle

\section{Introduction}\label{sec:intro}

The upcoming European Spallation Source (ESS)~\cite{Garoby:2017vew} facility is currently under construction on the outskirts of Lund, Sweden, and is expected to start operation in 2025.
With protons impinging on a fixed target, a large flux of neutrons are produced.
Meanwhile, a large number of light mesons including charged pions also arise, serving as the most intense source of neutrinos in the world.
This allows to probe various forms of new physics in the neutrino sector.
For instance, the future ESS Neutrino Super Beam (ESS$\nu$SB)~\cite{Alekou:2022emd} can be utilized for searches for light sterile neutrinos~\cite{Blennow:2014fqa,KumarAgarwalla:2019blx,Ghosh:2019zvl}.
To leverage this unique opportunity for searching for light new physics at the ESS, in Ref.~\cite{Soleti:2023hlr} a new experiment called Search for Hidden Neutrinos at the
ESS (SHiNESS) has been proposed, consisting of an active volume of 42 tons of liquid scintillator to be placed 25 meters away from the beam target.
Besides determining whether an eV-scale sterile neutrino can explain the short-baseline neutrino anomalies observed at various neutrino experiments in the past, and more precisely measuring the unitarity of active-neutrino mixing matrix, the SHiNESS experiment can search for long-lived MeV-scale heavy neutral leptons (HNLs) that mix with active neutrinos, and is projected to place new bounds with only two years of data acquisition.

In this work, we explore the capability of the SHiNESS experiment to probe long-lived particles (LLPs)~\footnote{See Refs.~\cite{Alimena:2019zri,Lee:2018pag,Curtin:2018mvb,Beacham:2019nyx} for recent reviews on LLP models and searches.} in additional theoretical scenarios of feebly interacting particles.
Indeed, in recent years after the observation of the Higgs boson at the LHC, lack of discovery of heavy fundamental particles that decay promptly as often predicted by new physics (NP) beyond the Standard Model (BSM) has led to increasingly stronger lower bounds on their masses and shifted more attention towards non-traditional forms of NP signatures.
These include displaced objects that are often associated with light LLPs.
For example, at the LHC, a list of ``far detectors'' have been proposed, dedicated to LLP searches, such as FASER~\cite{Feng:2017uoz,FASER:2018eoc,Feng:2022inv}, CODEX-b~\cite{Gligorov:2017nwh,Aielli:2019ivi}, and MATHUSLA~\cite{Curtin:2018mvb,Chou:2016lxi,MATHUSLA:2020uve}; in particular, FASER has been initiated since the start of Run 3 and published its early results~\cite{FASER:2023zcr,FASER:2023tle}.
These experiments have been shown to be particularly promising in searching for light LLPs in the MeV-GeV mass range predicted in a large class of BSM theories.
Here, we focus on the proposed SHiNESS experiment, where $\pi^+$ mesons are produced in large numbers and decay at rest.
We consequently study the experiment's potential in searching for light LLPs produced in decays of $\pi^+$ mesons and $\mu^+$ leptons which also originate from $\pi^+$ decays.

Concretely, we consider light binos in the R-parity-violating supersymmetry (RPV-SUSY)~\cite{Barbier:2004ez,Dreiner:1997uz,Mohapatra:2015fua}, and weak-violating electrophilic axionlike particles (ALPs)~\cite{Altmannshofer:2022ckw}, as benchmark models.
In the RPV-SUSY, the lightest neutralino with a GeV-scale or even smaller mass~\cite{Domingo:2022emr}, which is necessarily bino-like~\cite{Gogoladze:2002xp,Dreiner:2009ic}, is still allowed by all astrophysical and cosmological bounds~\cite{Grifols:1988fw,Ellis:1988aa,Lau:1993vf,Dreiner:2003wh,Dreiner:2013tja,Profumo:2008yg,Dreiner:2011fp} if we lift the GUT relation between the gauginos, $M_1\approx 0.5\, M_2$~\cite{Choudhury:1995pj,Choudhury:1999tn}, the light bino is unstable to avoid overclosing the Universe~\cite{Bechtle:2015nua}, and the dark matter does not consist of the lightest neutralino~\cite{Belanger:2002nr,Hooper:2002nq,Bottino:2002ry,Belanger:2003wb,AlbornozVasquez:2010nkq,Calibbi:2013poa}.
The lightest neutralino should also be the lightest supersymmetric particle.
The light bino can decay via RPV couplings, and if its mass is small, it has a long lifetime.
We note that \textit{a priori} the R-parity can be either conserved or broken, and therefore the RPV-SUSY is equally legitimate as the R-parity-conserving SUSY (RPC-SUSY)~\cite{Ibanez:1991pr,Dreiner:2005rd}.
Moreover, the RPV-SUSY offers a solution to the non-vanishing active-neutrino masses~\cite{Hall:1983id,Grossman:1998py,Hirsch:2000ef,Dreiner:2006xw,Dreiner:2011ft} and a richer phenomenology at colliders than the RPC-SUSY~\cite{Dreiner:1991pe,deCampos:2007bn,Dercks:2017lfq,Dreiner:2023bvs}.
It could also explain several experimental anomalies reported in recent years, including the $B$-anomalies~\cite{Trifinopoulos:2019lyo,Hu:2020yvs,BhupalDev:2021ipu}, the ANITA anomaly~\cite{Collins:2018jpg,Altmannshofer:2020axr}, as well as the muon $g-2$~\cite{Hu:2019ahp,Zheng:2021wnu,BhupalDev:2021ipu}.
The light binos in the RPV-SUSY can be produced in rare decays of, e.g.~bottom and charm mesons~\cite{deVries:2015mfw}.
Particularly, very few studies exist that constrain such long-lived light binos that are produced from decays of pions or muons, cf.~for instance, Refs.~\cite{Choudhury:1999tn,Dreiner:2023gir}.
Studying the sensitivity of the SHiNESS experiment to this theoretical scenario thus complements the existing efforts.
Recent phenomenological studies on long-lived light binos in the RPV-SUSY can be found in, e.g.~Refs.~\cite{deVries:2015mfw,Cottin:2022gmk,Gehrlein:2021hsk,Dercks:2018eua,Dreiner:2020qbi,Dreiner:2022swd,Gunther:2023vmz}.

ALPs could solve the strong CP problem~\cite{ParticleDataGroup:2022pth,Peccei:1977ur,Peccei:2006as} and in addition, serve as a dark-matter candidate~\cite{Dine:1982ah,Abbott:1982af,Preskill:1982cy,Marsh:2015xka,Lambiase:2018lhs,Auriol:2018ovo,Houston:2018vrf}, among other motivations.
They are often conceived as arising from the breaking of a global U(1) Peccei-Quinn (PQ)~\cite{Peccei:1977hh} symmetry as pseudo-Nambu-Goldstone bosons.
They are predicted to be long-lived for their feeble couplings with Standard-Model (SM) particles, even with masses in the MeV-GeV range.
The ALPs can couple to SM leptons, quarks, gauge bosons, etc.
Among these various possibilities, leptophilic ALPs are motivated for their connections to leptonic electric dipole moment~\cite{Kirpichnikov:2020lws}, charged lepton flavor violation~\cite{Han:2020dwo,Cheung:2021mol,Bertuzzo:2022fcm}, explaining the neutrino accesses at MiniBooNE~\cite{Chang:2021myh}, or accommodating the discrepancies observed between the measurements of the leptonic anomalous magnetic moment and the SM predictions~\cite{Bauer:2019gfk,Cornella:2019uxs,Buen-Abad:2021fwq,Bauer:2021mvw,Davoudiasl:2024vje}.
In this work, we focus on electrophilic ALPs with weak-violating couplings~\cite{Altmannshofer:2022ckw}.
In this case, a four-point vertex arises between an ALP, a $W$-boson, an electron, and an electron neutrino, with no helicity suppression present.
Some recent phenomenological works covering long-lived ALPs include Refs.~\cite{Bertholet:2021hjl,Co:2022bqq,Carmona:2022jid,Cheung:2022umw,Cheung:2021mol,Cheung:2024qve,Lu:2022zbe,Buonocore:2023kna}.

We emphasize that for the installation of the SHiNESS experiment, no upgrade to the beamline is required, and the experiment can start taking data as soon as the ESS launches its operation.
Moreover, as we will show, strong limits beyond current bounds on the considered models can be obtained in just a few years, strongly motivating the implementation of the SHiNESS experiment.

This paper is structured as follows.
In Sec.~\ref{sec:models} we introduce the theoretical models we study, and in Sec.~\ref{sec:experiment} we detail the setup of the SHiNESS experiment.
Sec.~\ref{sec:results} is devoted to presenting numerical results of the projected sensitivity reach of SHiNESS.
At the end, we summarize our findings in Sec.~\ref{sec:conclusions}.

\section{Theoretical models}\label{sec:models}

\subsection{Light binos in the RPV-SUSY}\label{subsec:bino}

In the RPV-SUSY, we have the following superpotential containing lepton-number-violating (LNV) and baryon-number-violating (BNV) terms, in addition to the usual R-parity-conserving terms:
\begin{eqnarray}
    W_\text{RPV}& = &\kappa_i L_i H_u + \frac{1}{2}\lambda_{ijk} L_i L_j \bar{E}_k  +  \lambda'_{ijk} L_i Q_j \bar{D}_k     \nonumber \\
    &&+  \frac{1}{2}\lambda''_{ijk}  \bar{U}_i \bar{D}_j \bar{D}_k,
\end{eqnarray}
where the operators in the first and the second lines break the lepton number and the baryon number, respectively, and $i, j, k = 1, 2, 3$ are generation indices.
$\kappa_i$ are couplings of mass dimension 1, and $\lambda$, $\lambda'$, and $\lambda''$ are dimensionless couplings.
Here, the factor $\frac{1}{2}$ is attached to the operators $LL\bar{E}$ and $\bar U \bar D \bar D$ because these operators possess anti-symmetry property such that $\lambda_{ijk} = - \lambda_{jik}$ and $\lambda''_{ijk} = - \lambda''_{ikj}$.
In this work, we assume that only $\lambda$ and $\lambda'$ couplings with certain flavor combinations are non-zero; this can be theoretically realized, as the bilinear term can be rotated away at a given energy scale~\cite{Hall:1983id,Dreiner:2003hw} and the BNV term can be vanishing if, for example, the baryon-triality symmetry is imposed~\cite{Dreiner:2005rd,Ibanez:1991hv,Martin:1997ns,Grossman:1998py,Dreiner:2006xw,Dreiner:2007uj,Bernhardt:2008jz}.

As discussed in Sec.~\ref{sec:intro}, a light MeV-scale bino is allowed and can be long-lived for tiny RPV couplings.
For the SHiNESS experiment, we focus on LLPs produced in $\pi^+$ and $\mu^+$ decays at rest and on the final-state signature of an electron-positron pair from the $\tilde{\chi}^0_1$ decays.
The $\pi^+$ decays into a light bino can be induced by the $\lambda'_{111}$ or $\lambda'_{211}$ couplings, while the signal decays of the $\mu^+$ lepton and the bino can be mediated by various $\lambda$ couplings.
For the light bino produced with the $\lambda'_{211}$ coupling in $\pi^+\to \mu^+ \tilde{\chi}^0_1$ decays, SHiNESS has no sensitivity, as we will explain later.
In Table~\ref{tab:bino_benchmarks} we list four RPV benchmark scenarios we choose to study.
In each of them, we assume only two RPV LNV couplings are non-vanishing, mediating production and decay of the lightest neutralino, respectively, except in one benchmark \textbf{B2} where only a single coupling $\lambda_{121}$ is assumed to be non-zero and it leads to both the production and decay of the light bino.
Specifically, in Table~\ref{tab:bino_benchmarks}, we list the production and decay couplings for the light bino, as well as the corresponding signal processes.
In addition, in some benchmark scenarios the production coupling can induce decays of the light bino in the interested mass range of the listed benchmark scenarios.
For instance, in \textbf{B1}, the decay $\tilde{\chi}^0_1 \to \pi^0 \overset{(-)}{\nu}_e$ can, in principle, take place via $\lambda'_{111}$, but the allowed bino mass window, $m_{\pi^0} < m_{\tilde{\chi}^0_1} < m_{\pi^+}$, is tiny, and therefore we ignore this case.
Also, in \textbf{B2}, by definition the production and decay couplings are identical.
Finally, in \textbf{B3}, the production coupling $\lambda_{121}$ can induce the $\tilde{\chi}^0_1\to \overset{(-)}{\nu}_\mu e^+ e^-$ decays which are taken into account in the numerical computation.

\begin{table}[t]
\resizebox{\textwidth}{!}{
\begin{tabular}{c|cccccc}
Benchmark & $\lambda_P$              & Prod.~via $\lambda_P$               & $\lambda_D$  & Decay via $\lambda_D$ &       Prod.~via $\lambda_D$  & Decay via $\lambda_P$                                   \\
\textbf{B1}   & $\lambda'_{111}$ &  $\pi^+\to e^+ \tilde{\chi}^0_1$    & $\lambda_{131}$ & $\tilde{\chi}^0_1\to \overset{(-)}{\nu}_\tau e^+ e^-$ & $-$ & $-$\\
\hline
\textbf{B2}   & $\lambda_{121}$ & $\mu^+\to \tilde{\chi}^0_1  \bar{\nu}_e e^+$ & $\lambda_{121}$ & $\tilde{\chi}^0_1\to \overset{(-)}{\nu}_\mu e^+ e^-$  & Redundant & Redundant  \\
\hline
\textbf{B3}   & $\lambda_{121}$ & $\mu^+\to \tilde{\chi}^0_1  \bar{\nu}_e e^+$ & $\lambda_{131}$ & $\tilde{\chi}^0_1\to \overset{(-)}{\nu}_\tau e^+ e^-$ &$-$  &$\tilde{\chi}^0_1\to \overset{(-)}{\nu}_\mu e^+ e^-$ \\
\hline
\textbf{B4} & $\lambda_{312}$ & $\mu^+\to \tilde{\chi}^0_1  \nu_\tau e^+$  & $\lambda_{131}$ & $\tilde{\chi}^0_1\to \overset{(-)}{\nu}_\tau e^+ e^-$ &  $-$ &$-$   \\               
\end{tabular}
}
\caption{Summary of the representative benchmark scenarios of the light binos in the RPV-SUSY that we study.
For benchmarks \textbf{B2-B4}, the $\mu^+$ leptons are produced in $\pi^+\to \mu^+ \nu_\mu$ decays at rest at the ESS and they also decay immediately at rest. The lightest neutralino is of Majorana nature and can hence decay to both charge-conjugated channels. {``$-$'' means either theoretically irrelevant, or kinematically forbidden.}}
\label{tab:bino_benchmarks}
\end{table}

To compute the production rates of the light bino from pion and muon decays via $\lambda'$ and $\lambda$ couplings, respectively, we resort to the analytic expressions provided in Refs.~\cite{deVries:2015mfw,Dreiner:2023gir}, while the decay widths' computation of the light bino via $\lambda$ couplings follows the procedure given in Ref.~\cite{Dreiner:2023gir}.

We now discuss the present bounds on the relevant RPV couplings.
These bounds stem from either measurements on low-energy observables~\cite{Allanach:1999ic,Dreiner:2023gir} or recast of past searches for long-lived HNLs~\cite{Dreiner:2023gir}.

For the coupling $\lambda'_{111}$, both measurements of neutrinoless double-beta ($0\nu\beta\beta$) decays and existing bounds on long-lived HNLs provide the leading limits on $\lambda'_{111}/m^2_{\tilde{f}}$ as functions of the lightest neutralino's mass $m_{\tilde{\chi}^0_1}$.
The bound from $0\nu\beta\beta$ decays on the coupling was obtained in Refs.~\cite{Mohapatra:1986su,Hirsch:1995zi,Hirsch:1995ek,Babu:1995vh}.
The reinterpretation of the HNL searches' results in terms of the light binos in the RPV-SUSY was performed in Ref.~\cite{Dreiner:2023gir}, and its Fig.~2 shows that these bounds are stronger than those from the $0\nu\beta\beta$ decays for $m_{\tilde{\chi}^0_1}$ roughly between 30 MeV and 130 MeV.

The involved $\lambda$ couplings all receive bounds from low-energy processes, as listed in Ref.~\cite{Allanach:1999ic}.
The bound on $\lambda_{121}$ originates from charged-current universality (C.~C.~universality)~\cite{Barger:1989rk}.
The upper limit on $\lambda_{131}$ was derived from measurements of $R_\tau=\Gamma(\tau\to e\nu \bar{\nu})/\Gamma(\tau\to \mu\nu\bar{\nu})$~\cite{Barger:1989rk}.
Moreover, the current bound on $\lambda_{312}$ stems from measurements on $R_{\tau\mu}=\Gamma(\tau\to \mu\nu\bar{\nu})/\Gamma(\mu\to e\nu\bar{\nu})$~\cite{Ledroit:1998}.

In addition, the $\lambda_{121}$ coupling is bounded from uncertainty on the muon decay width of $\mu^\pm \to e^\pm +$invisible~\cite{ParticleDataGroup:2022pth} ($\sigma_{\Gamma(\mu^\pm\to e^\pm+\text{invis.})}$) as functions of $m_{\tilde{\chi}^0_1}$, as given in Fig.~6 of Ref.~\cite{Dreiner:2023gir}.
The results are comparable with the bounds obtained from low-energy observables.

\begin{table}[t]
\resizebox{\textwidth}{!}{
\begin{tabular}{cccc}
Coupling                          & Relevant process & Resulting upper bound & References \\
\hline
\multirow{2}{*}{$\lambda'_{111}$} &     $0\nu\beta\beta$             &     $  5.2\cdot 10^{-4}\cdot \Big(\frac{m_{\tilde{e}}}{100\text{ GeV}}\Big)^2\cdot \sqrt{\frac{m_{\tilde{\chi}^0_1}}{100 \text{ GeV}}}$                 &  \cite{Mohapatra:1986su,Hirsch:1995zi,Hirsch:1995ek,Babu:1995vh}         \\
                                  &     HNL search recast             &  a function of $m_{\tilde{\chi}^0_1}$ and $m_{\tilde{f}}$                    &    \cite{Dreiner:2023gir}       \\
\hline
\multirow{2}{*}{$\lambda_{121}$}  &    C.~C.~universality              &     $0.049\cdot \frac{m_{\tilde{e}_R}}{100\text{ GeV}}$                 &  \cite{Barger:1989rk}         \\
                                  &    $\sigma_{\Gamma(\mu^\pm\to e^\pm+\text{invis.})}$              &      a function of $m_{\tilde{\chi}^0_1}$ and $m_{\tilde{f}}$                &      \cite{Dreiner:2023gir}     \\
\hline
$\lambda_{131}$                   &    $R_\tau$              &    $0.062\cdot \frac{m_{\tilde{e}_R}}{100\text{ GeV}}$                 &  \cite{Barger:1989rk}         \\
\hline
$\lambda_{312}$                    &     $R_{\tau\mu}$             &   $0.062\cdot \frac{m_{\tilde{\mu}_R}}{100\text{ GeV}}$                   &       \cite{Ledroit:1998}   
\end{tabular}
}
\caption{Summary of the current constraints on the RPV couplings involved in the considered benchmark scenarios listed in Table~\ref{tab:bino_benchmarks}.
The relevant processes, resulting bounds, as well as the references where these bounds were obtained, are tabulated.
For the $0\nu\beta\beta$ bound on $\lambda'_{111}, m_{\tilde{e}}$ labels the selectron's mass.
For the bounds on the $\lambda$ couplings, $\tilde{e}_R$ and $\tilde{\mu}_R$ denote right-handed selectron and smuon, respectively.
The bound on $\lambda'_{111}$ from HNL search recast and that on $\lambda_{121}$ from $\sigma_{\Gamma(\mu^\pm\to e^\pm+\text{invis.})}$ were derived in Ref.~\cite{Dreiner:2023gir} and given in the $\lambda'_{111}/m^2_{\tilde{f}}$ vs.~$m_{\tilde{\chi}^0_1}$ and $\lambda_{121}/m^2_{\tilde{f}}$ vs.~$m_{\tilde{\chi}^0_1}$ planes, respectively, with no analytic expression.
}
\label{tab:summary_rpv_bounds}
\end{table}
In Table~\ref{tab:summary_rpv_bounds}, we summarize the present bounds on these RPV couplings, including the relevant processes, resulting constraints, as well as the corresponding references.

In this work, we assume degenerate sfermion masses $m_{\tilde{f}}$ for simplicity.

Finally, it is worth noting that most of these existing bounds on the RPV couplings scale with $m_{\tilde{f}}$.
On the other hand, for the signal decay processes of the pion, muon, and light bino considered in this work, the decay widths are proportional to $\lambda^{(')2}/m^4_{\tilde{f}}$~\cite{deVries:2015mfw,Dreiner:2023gir}.
Therefore, these present bounds, when compared with the sensitivity reach of the SHiNESS experiment to this theoretical scenario as we will show later in Sec.~\ref{sec:results}, will be displayed for different sfermion masses.

\subsection{Electrophilic ALPs with weak-violating couplings}\label{subsec:alp}

We consider an ALP, $a$, coupled to electrons only, and study the following signal processes: $\pi^+ \to e^+ \, \nu_e \, a$ and $a\to e^+ e^-$, where $a$ can be long-lived for small couplings to the electrons.
If the ALP is radiated off from the electron in the charged-pion decays, the process is strongly suppressed by helicity flip.
However, as pointed out in Ref.~\cite{Altmannshofer:2022ckw}, a four-point vertex between an ALP, a $W$-boson, a charged lepton, and a neutrino, should arise for ALPs with weak-violating couplings with charged leptons and was previously missed in the literature.
The four-point vertex can result in large rates of $\pi^+\to e^+\, \nu_e\, a$ with no helicity-flip suppression.
Here, we focus on the following effective Lagrangian with a weak-violating term~\cite{Altmannshofer:2022ckw,Lu:2022zbe},
\begin{eqnarray}
    \mathcal{L}\supset \partial_\mu a \, \frac{c_{ee}}{2\Lambda}\bar{e} \gamma^\mu \gamma_5 e,
\end{eqnarray}
which, after integration by part is applied, is transformed to the following terms,
\begin{eqnarray}
    \mathcal{L} &\supset & i   \frac{c_{ee} \, m_e}{\Lambda} \,   a \, \bar{e}\, \gamma_5 \,  e   +  \frac{i g}{2\sqrt{2}}\frac{c_{ee}}{\Lambda}\,a \,  (\bar{e}\, \gamma^\mu  P_L \,  \nu_e) \, W_\mu^-  \nonumber \\
    &&+ \ldots + \text{h.c.}, 
    \label{eqn:Lagrangian_ALP}
\end{eqnarray}
where $g$ is the SU(2) coupling constant, $c_{ee}$ is a dimensionless coupling, and $m_e$ is the electron mass.
$\Lambda$ is the effective cut-off scale of new physics and $P_L$ is the left-chiral projector.
The dots denote a set of terms arising from chiral anomaly and irrelevant to the phenomenology in our study, and are therefore ignored here.

Both terms in Eq.~\eqref{eqn:Lagrangian_ALP} mediate the $\pi^+\to e^+ \, \nu_e \, a$ decay, and for the computation of its decay width, we refer to Ref.~\cite{Altmannshofer:2022ckw}.
The decay width of the ALP into an electron-positron pair is calculated with the following expression~\cite{Bauer:2017ris,Bauer:2018uxu,Lu:2022zbe}
\begin{eqnarray}
    \Gamma(a\to e^+ e^-) = \frac{c_{ee}^2}{8\pi \Lambda^2} m_e^2 m_a \sqrt{1 - \frac{4 m^2_{e}}{m_a^2}},
\end{eqnarray}
where $m_a$ is the ALP mass.
Following the practice taken in Ref.~\cite{Altmannshofer:2022ckw}, we assume that the ALP decays with a 100\% branching ratio into an electron-positron pair, and ignore the loop-induced photon-pair decay mode (which is sub-dominant for the relevant mass range anyway).
It is worth noting that both the $\pi^+\to e^+ \, \nu_e \, a$ and $a\to e^+ e^-$ decays are mediated by the $c_{ee}$ coupling.

The existing bounds on $c_{ee}/\Lambda$ in the weak-violating scenario are obtained from various laboratory and astrophysical searches and observations~\cite{BaBar:2014zli,Konaka:1986cb,Riordan:1987aw,Bjorken:1988as,Bross:1989mp,Morel:2020dww,Parker:2018vye,Lucente:2021hbp,KTeV:2003sls,LHCb:2015ycz,NA62:2020xlg,SINDRUM:1986klz,Poblaguev:2002ug,CHARM:1985anb}.
In addition, rough potential projections of sensitivities, expected at the PIONEER experiment~\cite{PIONEER:2022yag}, kaon factories~\cite{Goudzovski:2022vbt}, as well as the LHC, are provided in Ref.~\cite{Altmannshofer:2022ckw}.
In particular, the latter corresponds to a benchmark value of $10^{-5}$ for the ratio $\text{Br}_W=\text{Br}(W^+\to l^+ \nu_l a)/\text{Br}(W^+\to e^+ \nu)$, expected at the future LHC.

At the end, we comment that Ref.~\cite{Altmannshofer:2022ckw} also studied a weak-preserving scenario where the contributions from the four-point vertex are vanishing.
We explicitly checked and found that in this case the expected sensitivity reach of SHiNESS is already all excluded by the SLAC E137 experiment~\cite{Bjorken:1988as}, if we restrict ourselves to the ALP coupled to electrons only, as the strong helicity suppression largely reduces the productions rates of the ALP.
If the ALP is coupled to muons only, it will be produced in $\pi^+\to \mu^+ \nu_\mu a$ decays; in this case, the helicity suppression is negligible, but SHiNESS will still have no sensitivity, for kinematic reasons related to cut selections to be imposed on final-state leptons (see discussion in Sec.~\ref{sec:experiment}).
Therefore, we do not include weak-preserving scenarios in the present study.

\section{Experimental setup}\label{sec:experiment}

At the upcoming ESS facility, a 2-GeV proton beam hits a rotating tungsten target, producing large numbers of neutrons, serving the facility's main purpose~\cite{Garoby:2017vew}.
In total, $2.8\times 10^{23}$ protons on target are planned for each calendar year.
In such proton-nucleus collisions, an enormous number of charged pions among other hadrons are produced.
While negatively charged pions are absorbed by the nucleus before decaying, positively charged ones lose energy {via ionization effects} while they are traveling and finally decay at rest, leading to the following decay chain:
\begin{eqnarray}
    \pi^+ \to \mu^+ \nu_{\mu}, \,\,\, \mu^+ \to e^+ \nu_e \, \bar{\nu}_\mu.
\end{eqnarray}
Simulation predicts an expected annual rate of $8.5 \times 10^{22}$ decays at rest for both $\pi^+$ mesons and $\mu^+$ leptons at the ESS~\cite{Baxter:2019mcx}.

The SHiNESS experiment has been proposed to be installed at some available space inside the D03 experimental hall, with a distance of 25 meters away from the tungsten target.
Further, the SHiNESS detector is supposed to be placed with a polar angle $\theta=35^\circ$ in the backward direction of the incoming proton beam, for the purpose of background-event mitigation.
The detector is planned to be a stainless-steel cylindrical tank of a 5.3-m height and a 3.3-m radius, containing an internal active volume of a 4-m height and a 2-m radius filled with loaded liquid scintillator.
In this work, we treat the active volume inside as the fiducial volume of the detector.

The proposal of the SHiNESS detector focuses on the $e^- e^+$-pair signature in the context of searches for LLPs such as long-lived HNLs.
This aim is realized by installing arrays of photomultipliers.
Further, with the ability to detect and distinguish between both Cherenkov photons and scintillation light, the SHiNESS detector has strong vertex resolution and can reconstruct particle directions.
In particular, the separation of the Cherenkov and scintillation light is allowed by various strategies and hardware implementation such as photosensors for their excellent timing resolution.
Thus, the electron and the positron in a signal event can be separated, and the energy resolution of the charged leptons is at the $\mathcal{O}(1)\%$ level, cf.~Fig.~8 of Ref.~\cite{Soleti:2023hlr}.
We refer to Ref.~\cite{Soleti:2023hlr} for more detail.
In this work, as already specified in the previous section, we also concentrate on the signature of excesses in the number of $e^- e^+$ events (coming from decays of long-lived binos or ALPs).

We follow Ref.~\cite{Soleti:2023hlr} for a brief discussion on background estimates.
The LLP search requires reconstruction of a vertex formed from two charged leptons.
For this purpose, the two lepton signals should be separated in time within 100 ns, and both should have an energy larger than 17 MeV.
In addition, the opening angle between the two leptons should be at least $15^\circ$~\footnote{The lepton cuts imply that the light bino produced in $\pi^+\to \mu^+ \tilde{\chi}^0_1$ decays at rest via the coupling $\lambda'_{211}$ cannot be probed by SHiNESS because the light bino would have at most an energy of $m_{\pi^+}-m_{\mu^+}\approx 35$ MeV.
As a result, it is impossible for the electron-positron pair produced in the decays of the light bino to pass the lepton-cut selections required in the search.
For the same reason, SHiNESS is insensitive to the ALP produced in $\pi^+\to \mu^+ \, \nu_\mu \, a$ decays.}.
With these requirements, the main background sources are cosmic ray, charged-current $\nu_e$ interactions, neutral-current $\nu$ interactions, and $\bar{\nu}_e$ beam component.
These sources sum up to about 61 background events per year.
A large, $50\%$ systematic uncertainty is assigned, corresponding to 31 events per year~\cite{Soleti:2023hlr}, and we take the same assumption in this work.
For a benchmark 2-year operation time, we thus expect $122\pm 43.84$ background events in total, corresponding to an NP signal-event number of 72.17 for exclusion bounds at 90\% confidence level (CL).

Using Monte-Carlo techniques, we estimate the efficiency of the LLP decay products fulfilling a specified list of criteria.
For each simulated event, we check first on the final-state electron and positron whether they pass the above-mentioned cuts.
If the energy and opening-angle requirements are satisfied, the associated simulated LLP is taken for computing its decay probability inside the active volume of the SHiNESS detector.
The probability of the $i^{\text{th}}$ simulated LLP to pass the lepton cuts and to decay inside the fiducial volume of SHiNESS is calculated as follows
\begin{eqnarray}
	\epsilon^{\text{cut\&acc.}}_{i^\text{th}\text{ LLP}} =
\begin{cases}
0, \text{ if failing the lepton cuts,}\\
\epsilon^{\text{f.v.}}_{i^\text{th}\text{ LLP}}, \text{ if passing these selections,}
\end{cases}
\end{eqnarray}
where we include the cut requirements on the final-state electron and positron, ``f.v.'' in $\epsilon^{\text{f.v.}}_{i^\text{th}\text{ LLP}}$ stands for ``fiducial volume'', and ``acc.'' is abbreviation for ``acceptance''.
To compute $\epsilon^{\text{f.v.}}_{i^\text{th}\text{ LLP}}$, we assume that the target is point-like and located in the middle of the SHiNESS detector in the vertical dimension.
\begin{figure}[t]
	\centering
	\includegraphics[width=0.7\textwidth]{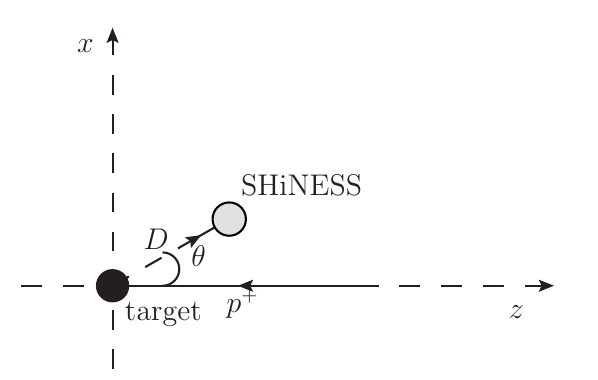}
	\caption{A bird's-eye view of the SHiNESS experiment. $\theta=35^\circ$, and $p^+$ denotes the incoming 2-GeV proton beam. The distance of SHiNESS from the target is 25 meters. $x$ and $z$ label the axes.
 }
 \label{fig:detector_sketch}
\end{figure}
In Fig.~\ref{fig:detector_sketch} we display a bird's-eye view of the experimental setup, where $\theta=35^\circ$ and $D=25$ m is the distance of the detector to the tungsten target. 
The center of SHiNESS is at $y=0$.
With the Monte-Carlo simulations, we calculate the decay-in-volume probability, $\epsilon^{\text{f.v.}}_{i^\text{th}\text{ LLP}}$, by taking into account the kinematics and the lifetime of the $i^{\text{th}}$ simulated LLP, as well as the previously defined fiducial volume,  if it has satisfied the lepton-cut requirements.

The signal-event number $N_S$ is computed with the following formula,
\begin{eqnarray}
	N_S= N_{\text{LLP}}\cdot  \epsilon^{\text{cut\&acc.}}_{\text{avg.}}\cdot \text{Br}(\text{LLP}\to e^- e^+ X),
\end{eqnarray}
where $N_{\text{LLP}}$ is the total number of the LLPs expected to be produced in two years' operation of the ESS, $\text{Br}(\text{LLP}\to e^- e^+ X)$ is the decay branching ratio of the LLP into an electron-positron pair plus anything, and the average efficiency $\epsilon^{\text{cut\& acc.}}_{\text{avg.}}$ is evaluated with the following expression:
\begin{eqnarray}
	\epsilon^{\text{cut\&acc.}}_{\text{avg.}}=\frac{1}{N_{\text{sim.}}}\sum_{i=1}^{N_{\text{sim.}}} \epsilon^{\text{cut\&acc.}}_{i^\text{th}\text{ LLP}} \, .
\end{eqnarray}
Here, $N_{\text{sim.}}$ is the number of simulated signal events.

\begin{figure}[t]
	\centering
 \includegraphics[width=0.7\textwidth]{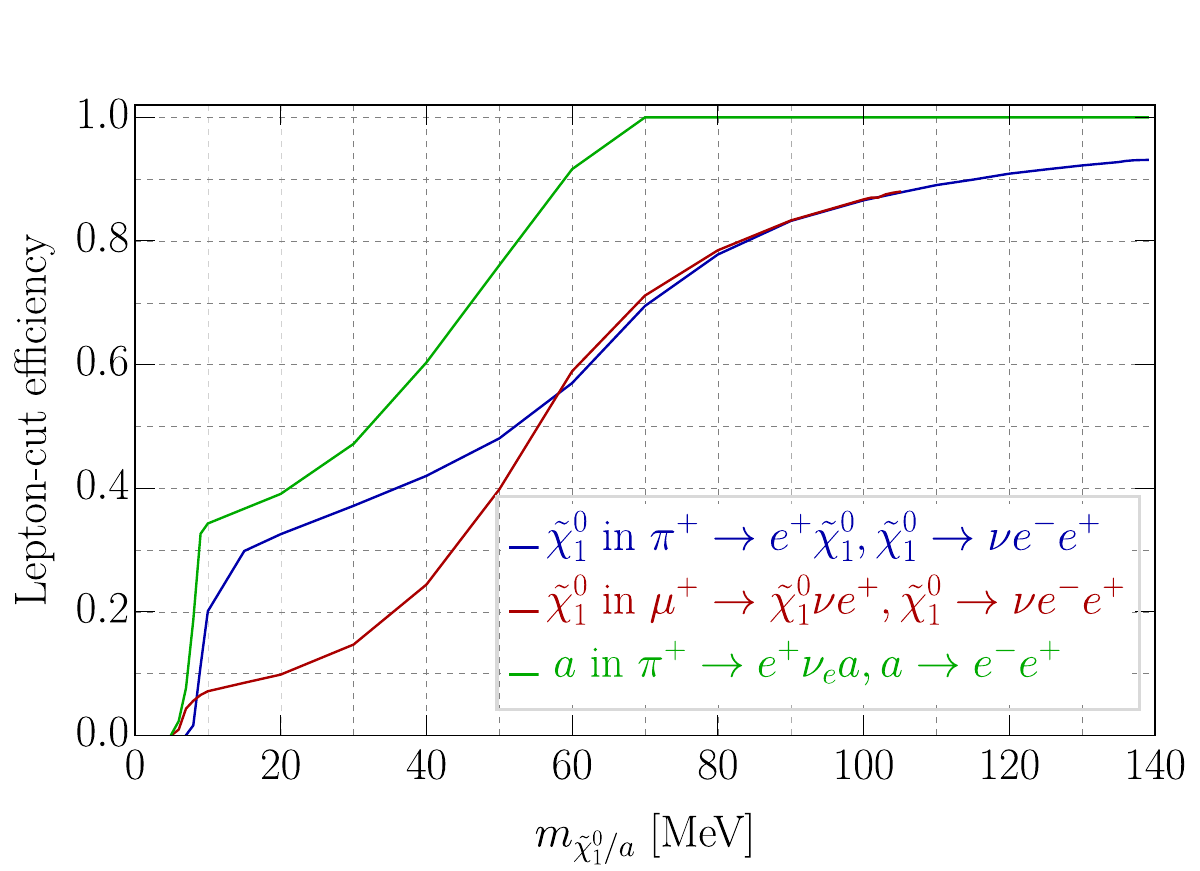}
	\caption{Lepton-cut efficiencies as functions of the mass of the bino or ALP, where only the requirements on the energies of the final-state electron and positron, as well as the opening angle between them, are included.
 The blue (scarlet) curve is for the bino benchmark \textbf{B1} (\textbf{B2-B4}), and the green one is for the electrophilic ALP model.
 }
 \label{fig:lepton_cut_efficiency}
\end{figure}
In Fig.~\ref{fig:lepton_cut_efficiency}, we show a plot for the average efficiency of the lepton cuts as functions of the LLP mass, where the fiducial-volume requirement is not included.
The blue and scarlet curves correspond to the bino benchmarks \textbf{B1} and \textbf{B2-B4}, respectively, and the green one is for the electrophilic ALP scenario.
The latter scenario has the strongest lepton-cut efficiencies, mainly because the ALP undergoes two-body decays into an electron-positron pair, while the light binos decay into three particles among which the active neutrino carries away energy resulting in less energetic electron and positron.
We find that the lepton-cut efficiencies lie in $\mathcal{O}(10)\%$ for almost the whole range of the kinematically allowed LLP mass.
However, the total efficiency $\epsilon^{\text{cut\&acc.}}_{\text{avg.}}$ depends also on the lifetime of the LLP since it includes the fiducial-volume requirement.
In our numerical simulation, we observe that the maximal value of $\epsilon^{\text{cut\&acc.}}_{\text{avg.}}$ is about $10^{-4}$ in all theoretical scenarios.

\section{Numerical results}\label{sec:results}

We present numerical results in this section, for the exclusion bounds at 90\% CL.

\subsection{Light binos}\label{subsec:results_bino}

For the theoretical model of long-lived light neutralinos in the RPV-SUSY, SHiNESS can probe large, unexcluded parameter space in the benchmark scenarios listed in Table~\ref{tab:bino_benchmarks}.

\begin{figure}[t]
	\centering
	\includegraphics[width=0.495\textwidth]{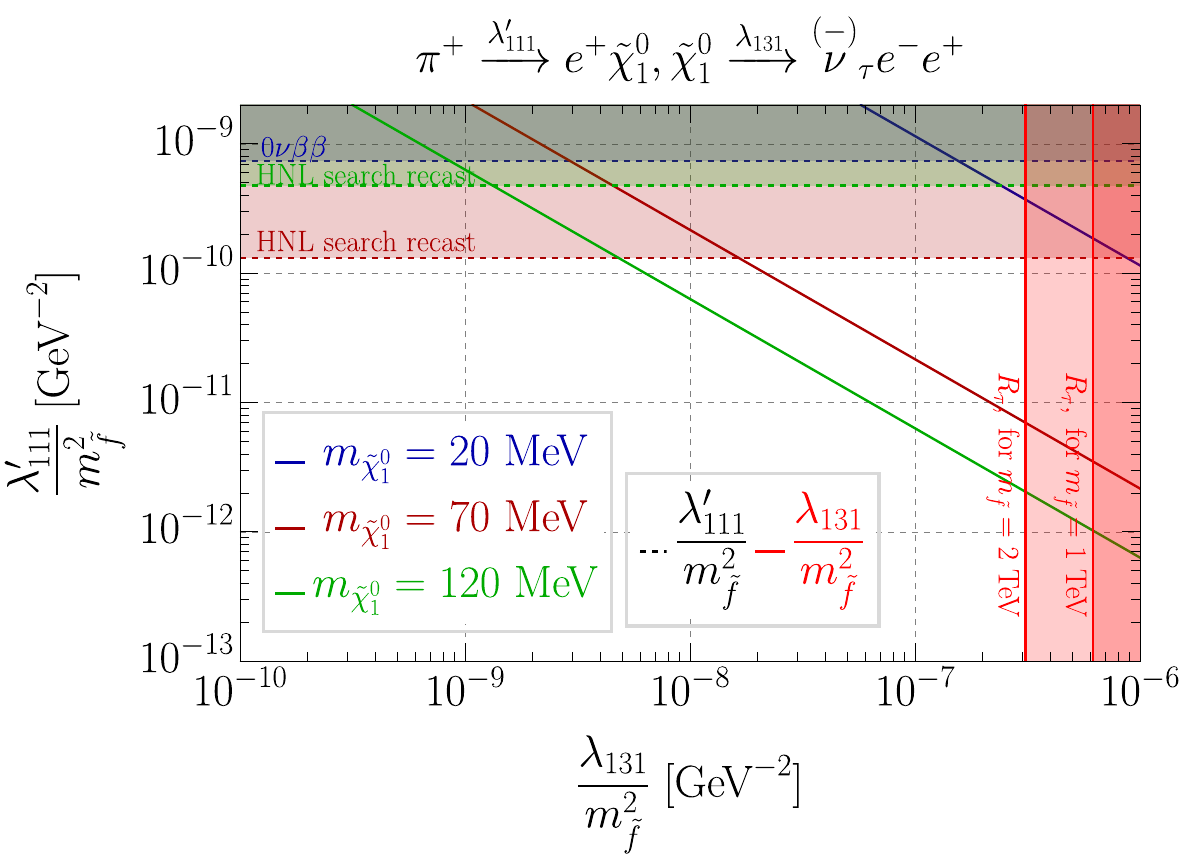}
 	\includegraphics[width=0.495\textwidth]{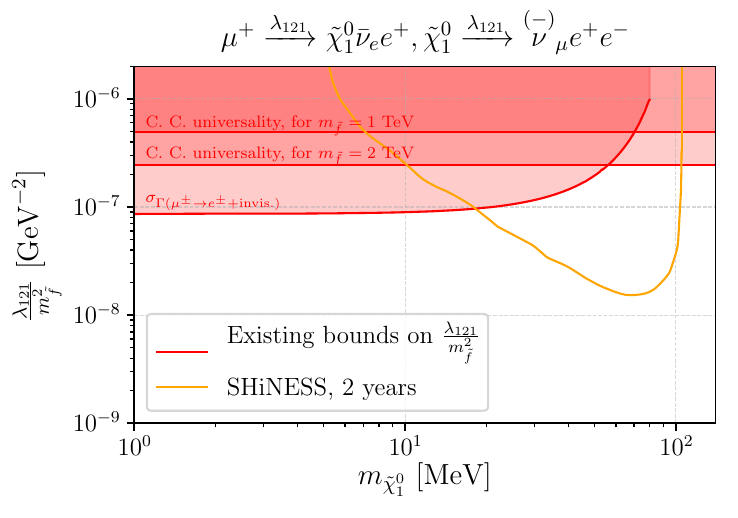}
	\caption{\textit{Left:} sensitivity reach of SHiNESS at 90\% CL in two years' data-collection time to the benchmark scenario \textbf{B1}, shown in the $(\lambda_{131}/m^2_{\tilde{f}}, \lambda'_{111}/m^2_{\tilde{f}})$ plane, for $m_{\tilde{\chi}^0_1}=20$ MeV, 70 MeV, and 120 MeV.
 The horizontal dashed curves are present bounds on $\lambda'_{111}/m^2_{\tilde{f}}$ obtained from $0\nu\beta\beta$ decays~\cite{Mohapatra:1986su,Hirsch:1995zi,Hirsch:1995ek,Babu:1995vh} and recasting HNL searches~\cite{Dreiner:2023gir} with the excluded regions shaded with the colors of the corresponding masses of $\tilde{\chi}^0_1$, and the vertical solid red curves are current limits on $\lambda_{131}/m^2_{\tilde{f}}$ for sfermion masses of 1 TeV and 2 TeV obtained by measuring $R_\tau$, with the area to the right of these vertical lines being already excluded, cf.~Table~\ref{tab:summary_rpv_bounds}.
 \textit{Right:} sensitivity reach of SHiNESS to the benchmark \textbf{B2}, displayed in the $\lambda_{121}/m^2_{\tilde{f}}$ vs.~$m_{\tilde{\chi}^0_1}$ plane. The red solid horizontal curves are current limits on $\lambda_{121}/m^2_{\tilde{f}}$ obtained by considering C.~C.~universality, cf.~Table~\ref{tab:summary_rpv_bounds}, for $m_{\tilde{f}}=1$ TeV and 2 TeV, and the third red solid curve is the bound obtained by considering $\sigma_{\Gamma(\mu^\pm\to e^\pm+\text{invis.})}$.
 }
 \label{fig:sensitivity_bino_prime111_or_131_or_121}
\end{figure}

For benchmark \textbf{B1}, strong bounds exist on the single coupling $\lambda'_{111}/m^2_{\tilde{f}}$ as functions of the light bino's mass, from $0\nu\beta\beta$ decays and HNL searches, cf.~the discussion in Sec.~\ref{sec:models}.
We present a sensitivity plot for \textbf{B1} in the left panel of Fig.~\ref{fig:sensitivity_bino_prime111_or_131_or_121}, shown in the $\lambda'_{111}/m^2_{\tilde{f}}$ vs.~$\lambda_{131}/m^2_{\tilde{f}}$ plane, where we fix the light bino's mass at 20 MeV (blue), 70 MeV (scarlet), and 120 MeV (green).
The horizontal dashed curves are current upper bounds on $\lambda'_{111}/m^2_{\tilde{f}}$, where we have picked, for each bino-mass choice, the stronger limit between that derived from $0\nu\beta\beta$ decays and that from HNL searches.
The vertical solid red curves are present upper limits on $\lambda_{131}/m^2_{\tilde{f}}$ for $m_{\tilde{f}}=1$ TeV and 2 TeV derived from measurements on $R_\tau$ (see Table~\ref{tab:summary_rpv_bounds}) which do not depend on $m_{\tilde{\chi}^0_1}$.
These present limits are labeled with the processes from which they were obtained.
We find that for all the three choices of the bino mass, a triangle-shaped window that is allowed in the shown parameter space can be excluded by SHiNESS in two years.

The benchmark \textbf{B2} is the only single-coupling scenario listed in Table~\ref{tab:bino_benchmarks}, where the only non-vanishing coupling $\lambda_{121}$ mediates both the production and decay of the lightest neutralino.
We show a sensitivity plot in the right panel of Fig.~\ref{fig:sensitivity_bino_prime111_or_131_or_121} in the $(m_{\tilde{\chi}^0_1}, \lambda_{121}/m^2_{\tilde{f}})$ plane.
Here, SHiNESS in two year's operation can probe $m_{\tilde{\chi}^0_1}$ up to right below the muon threshold.
Compared to the existing bounds from C.~C.~universality~\cite{Barger:1989rk} (red horizontal curves), and those from the uncertainty on the measured decay width of $\mu^\pm\to e^\pm+\text{invisible}$~\cite{Dreiner:2023gir} (the third red curve), we find that SHiNESS can exclude large parameter space for $m_{\tilde{\chi}^0_1}$ roughly between 20 MeV and 100 MeV, probing $\lambda_{121}/m^2_{\tilde{f}}$ down to as low as about $1.5\times 10^{-8}$ GeV$^{-2}$.

\textbf{B3} is the only benchmark (besides \textbf{B2}) where the production coupling also induces kinematically allowed bino decays.
\begin{figure}[t]
	\centering
 \includegraphics[width=0.495\textwidth]{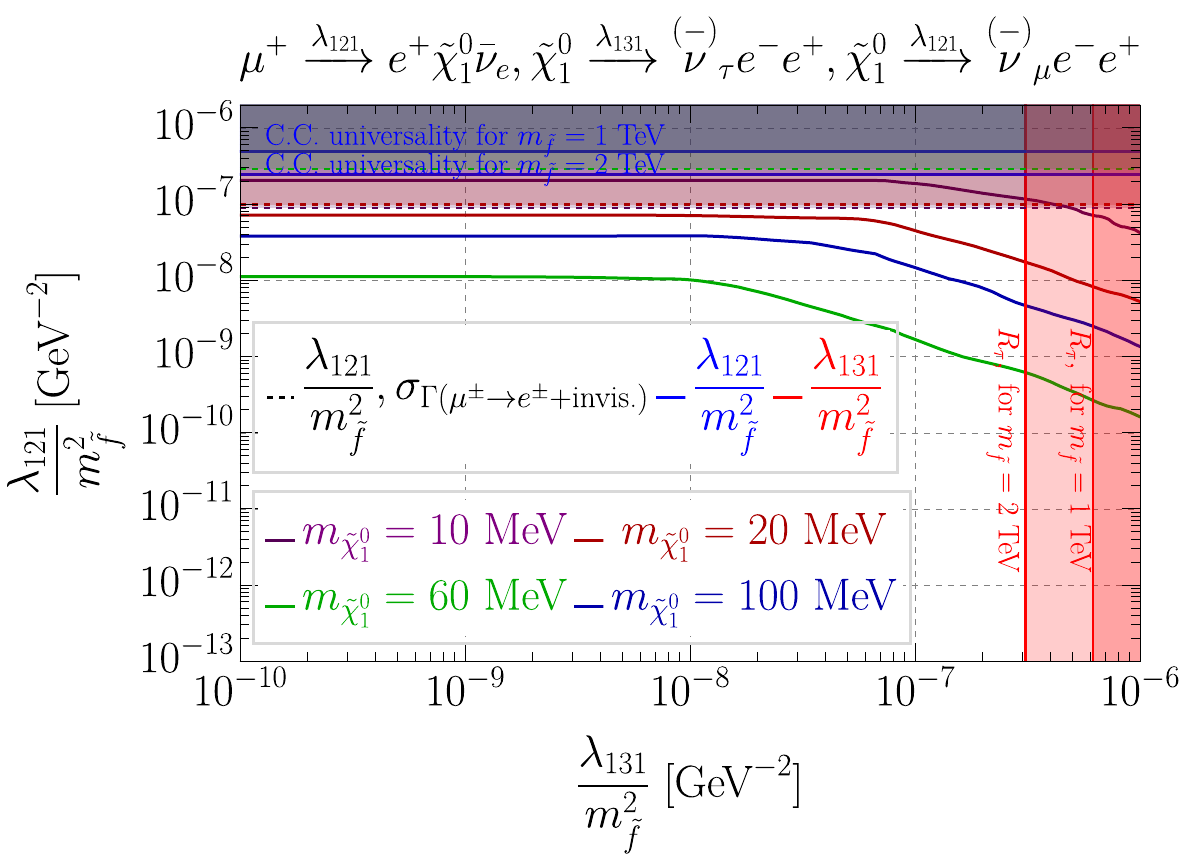}
 \includegraphics[width=0.495\textwidth]{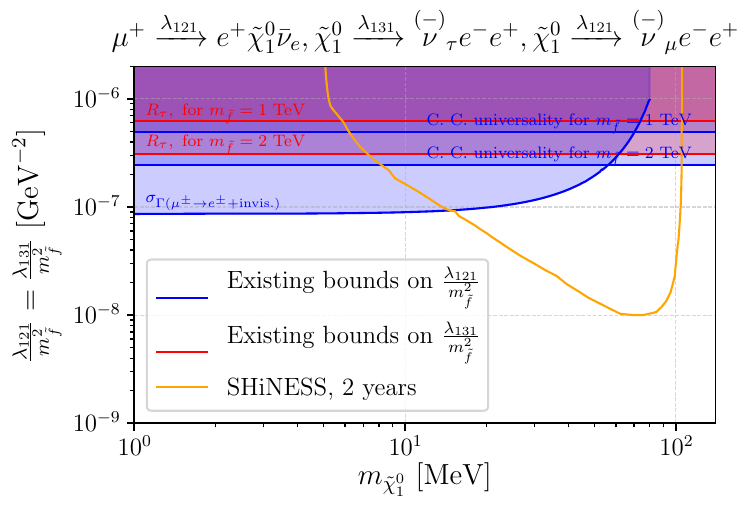}
	\caption{Sensitivity reach of SHiNESS to the benchmark \textbf{B3} presented in the $\lambda_{121}/m^2_{\tilde{f}}$ vs.~$\lambda_{131}/m^2_{\tilde{f}}$ plane (left plot) and the $\lambda_{121}/m^2_{\tilde{f}}=\lambda_{131}/m^2_{\tilde{f}}$ vs.~$m_{\tilde{\chi}^0_1}$ plane (right plot). 
	In the left panel, the bounds from $\sigma_{\Gamma(\mu^\pm\to e^\pm+\text{invis.})}$ on $\lambda_{121}/m^2_{\tilde{f}}$ for $m_{\tilde{\chi}^0_1}=100$ MeV are unavailable in Ref.~\cite{Dreiner:2023gir}.
 }
 \label{fig:sensitivity_bino_121_131}
\end{figure}
We choose to present the numerical results in Fig.~\ref{fig:sensitivity_bino_121_131}, with two plots shown in the $\lambda_{121}/m^2_{\tilde{f}}$ vs.~$\lambda_{131}/m^2_{\tilde{f}}$ and $\lambda_{121}/m^2_{\tilde{f}}=\lambda_{131}/m^2_{\tilde{f}}$ vs.~$m_{\tilde{\chi}^0_1}$ planes, respectively.
The present limits on both $\lambda_{121}/m^2_{\tilde{f}}$ and $\lambda_{131}/m^2_{\tilde{f}}$, derived from C.~C.~universality and $R_\tau$, are overlaid with blue and red curves, respectively, for both sfermion masses of 1 TeV and 2 TeV.
Moreover, for $\lambda_{121}/m^2_{\tilde{f}}$, the bounds obtained by considering $\sigma_{\Gamma(\mu^\pm\to e^\pm+\text{invis.})}$ are also shown, in dashed horizontal lines with different colors corresponding to the chosen mass values (10 MeV, 20 MeV, and 60 MeV) of the light bino (left panel) and in a blue solid line (right panel).
In the left plot of Fig.~\ref{fig:sensitivity_bino_121_131}, we observe that for small values of $\lambda_{131}/m^2_{\tilde{f}}$ the sensitivity curves become flat, reflecting the fact that the $\lambda_{121}$ coupling alone can lead to sufficiently many signal events.
We find that for $m_{\tilde{\chi}^0_1}=20$ MeV, 60 MeV, and 100 MeV, SHiNESS can probe large parameter space beyond the present bounds.
Further, for increasing $m_{\tilde{\chi}^0_1}$ the sensitivity of SHiNESS is enhanced, except for the $m_{\tilde{\chi}^0_1}=100$ MeV case where the phase-space suppression effect comes into play.
The right plot of Fig.~\ref{fig:sensitivity_bino_121_131}, similar to the right plot of Fig.~\ref{fig:sensitivity_bino_prime111_or_131_or_121}, shows that SHiNESS can extend the current bounds largely in just two years, mainly for $m_{\tilde{\chi}^0_1}$ between about 20 MeV and 100 MeV.

We present numerical results of \textbf{B4} next, where there is no available bound from reinterpretation of the HNL searches or $\sigma_{\Gamma(\mu^\pm\to e^\pm+\text{invis.})}$.
\begin{figure}[t]
	\centering
 \includegraphics[width=0.495\textwidth]{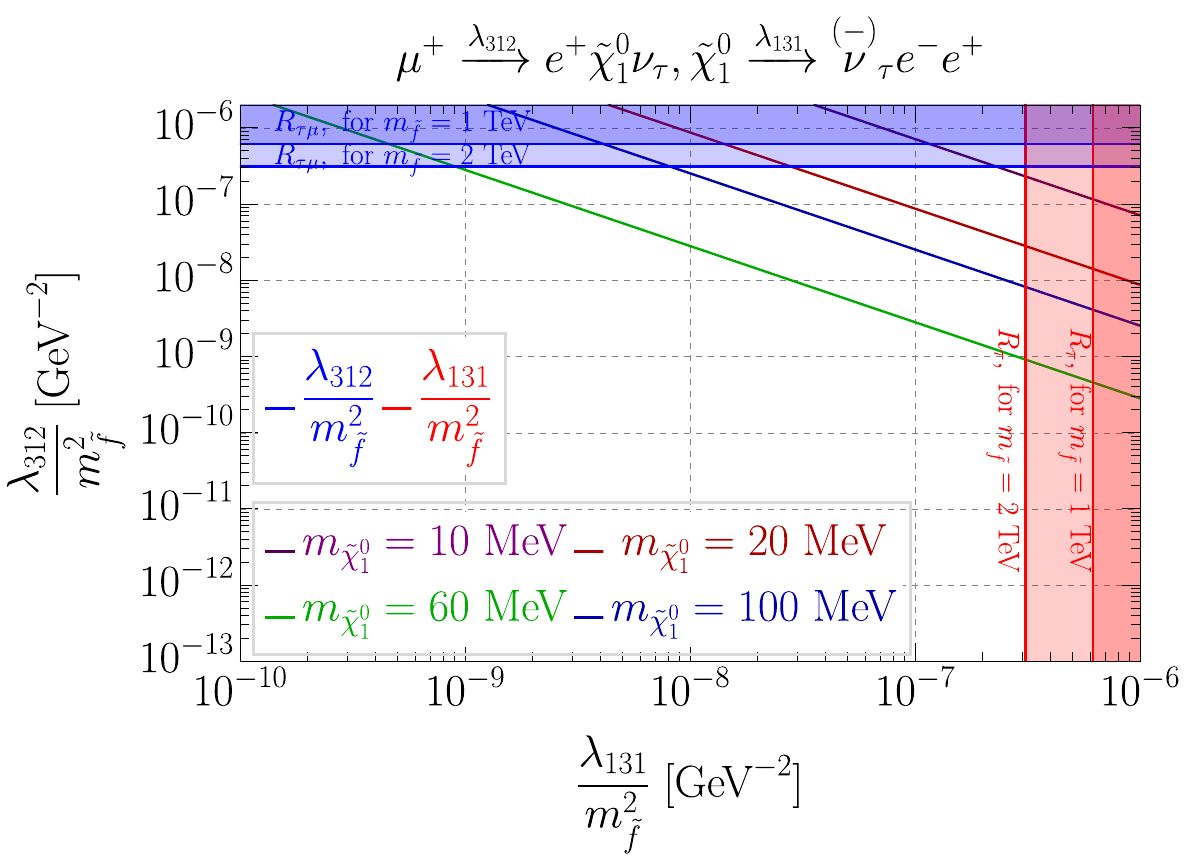}
 \includegraphics[width=0.495\textwidth]{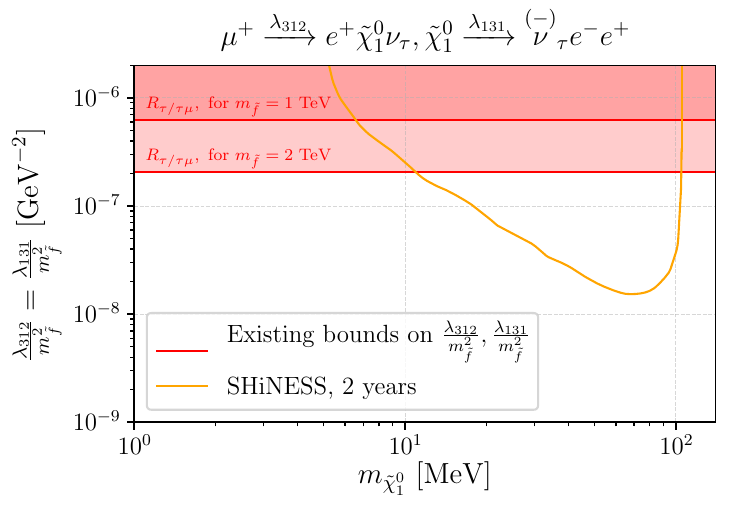}
	\caption{Sensitivity reach of SHiNESS to the benchmark \textbf{B4}, shown in the $\lambda_{312}/m^2_{\tilde{f}}$ vs.~$\lambda_{131}/m^2_{\tilde{f}}$ plane (left plot) and the $\lambda_{312}/m^2_{\tilde{f}}=\lambda_{131}/m^2_{\tilde{f}}$ vs.~$m_{\tilde{\chi}^0_1}$ plane (right plot).
 }
 \label{fig:sensitivity_bino_312_131}
\end{figure}
In Fig.~\ref{fig:sensitivity_bino_312_131}, we show two sensitivity plots in the $\lambda_{312}/m^2_{\tilde{f}}$ vs.~$\lambda_{131}/m^2_{\tilde{f}}$ and $\lambda_{312}/m^2_{\tilde{f}}=\lambda_{131}/m^2_{\tilde{f}}$ vs.~$m_{\tilde{\chi}^0_1}$ planes, respectively.
In the left plot, we fix the light neutralino's mass at 10 MeV, 20 MeV, 60 MeV, and 100 MeV.
The present bounds on the RPV couplings for $m_{\tilde{f}}=1$ TeV and 2 TeV are shown, and we find that in the optimal case ($m_{\tilde{\chi}^0_1}=60$ MeV), new parameter space up to more than two orders of magnitude beyond the current bounds can be excluded by SHiNESS with two years' data collection.
The right plot of Fig.~\ref{fig:sensitivity_bino_312_131} indicates that SHiNESS is predicted to probe unexcluded values of $\lambda_{312}/m^2_{\tilde{f}}=\lambda_{131}/m^2_{\tilde{f}}$ more than one order of magnitude beyond the present limits in two years' operation, for $m_{\tilde{\chi}^0_1}$ between 10 MeV and 100 MeV.

\subsection{Weak-violating ALPs coupled to electrons}\label{subsec:results_alp}

\begin{figure}[t]
	\centering
 \includegraphics[width=0.7\textwidth]{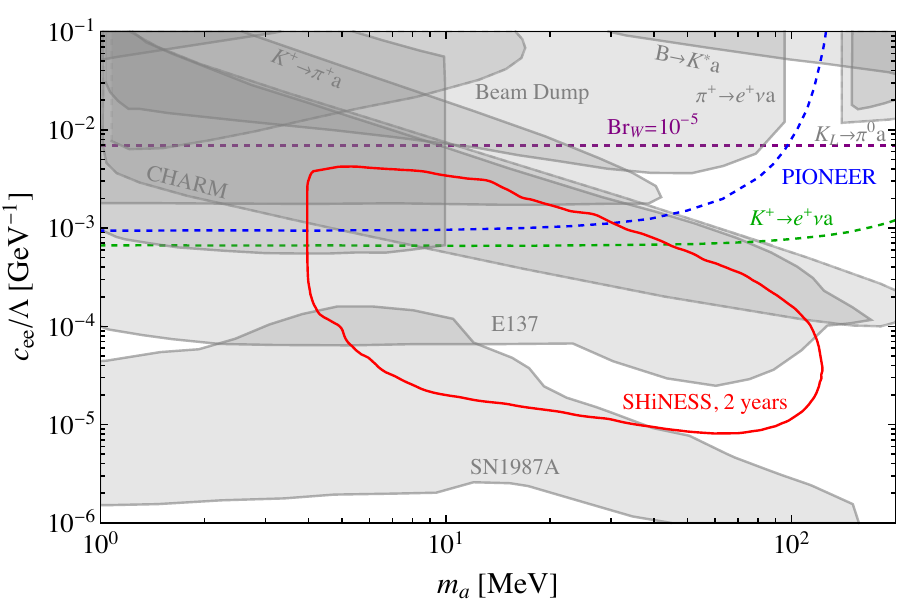}
	\caption{The projected sensitivity of SHiNESS with two years' operation to the weak-violating electrophilic ALP, displayed in the $(m_a, c_{ee}/\Lambda)$ plane.
 The area inside the red curve can be excluded by SHiNESS.
 The gray area in the background is the currently excluded parameter space, reproduced from Ref.~\cite{Altmannshofer:2022ckw}; see the text for detail.
 We also reproduce from Ref.~\cite{Altmannshofer:2022ckw} estimates of potential sensitivities from the PIONEER experiment~\cite{PIONEER:2022yag} (dashed blue), kaon factories~\cite{Goudzovski:2022vbt} (dashed green), as well as a benchmark of the LHC (dashed purple)~\cite{Altmannshofer:2022ckw}.
 See also the discussion in Sec.~\ref{subsec:alp}.
 }
 \label{fig:sensitivity_ALP_coupling_vs_mass}
\end{figure}

The projected sensitivities of SHiNESS with two-year data collation are presented in Fig.~\ref{fig:sensitivity_ALP_coupling_vs_mass}, in the plane $c_{ee}/\Lambda$ vs.~$m_a$.
The current bounds stemming from electron beam-dump experiments~\cite{Konaka:1986cb,Riordan:1987aw,Bjorken:1988as,Bross:1989mp}, supernova SN1987A~\cite{Lucente:2021hbp}, and searches for flavor-changing-neutral-current decays of mesons~\cite{KTeV:2003sls,LHCb:2015ycz,NA62:2020xlg}, are shown in gray.
Further, bounds obtained in searches for leptonic decays of charged mesons are included in the gray area, taking into account $\pi^+$ decays~\cite{SINDRUM:1986klz}, $K^+$ decays~\cite{Poblaguev:2002ug}, as well as the CHARM experiment~\cite{CHARM:1985anb}.
Particularly, rough estimates for potential bounds obtained from charged-pion and kaon decays are also overlaid with dashed lines, for the PIONEER experiment~\cite{PIONEER:2022yag} (blue), kaon factories~\cite{Goudzovski:2022vbt} (green), and the LHC (purple)~\cite{Altmannshofer:2022ckw}.
We observe that while these future experiments are estimated to probe mainly the unexcluded parameter region of $c_{ee}/\Lambda \gtrsim 6\times 10^{-4}$ GeV$^{-1}$, SHiNESS can be sensitive to a unique, currently allowed region $c_{ee}/\Lambda \sim \mathcal{O}(10^{-5})$ GeV$^{-1}$ for $m_a$ between roughly 10 MeV and 120 MeV.

\section{Conclusions}\label{sec:conclusions}

In this work, we have investigated the potential of the SHiNESS experiment proposed to be installed at the ESS facility, to search for certain light long-lived particles.
Following the SHiNESS proposal~\cite{Soleti:2023hlr}, we focus on the signature of LLPs at SHiNESS as excesses in electron-positron pair events, and selection cuts on the final-state leptons are imposed in order to suppress background events to an acceptable level.
Moreover, in order to estimate the cut efficiencies and acceptance of SHiNESS to LLPs, we have performed numerical simulations.

The focus of this work is on LLPs produced in rare decays at rest of $\pi^+$ mesons that arise from the 2-GeV proton beam at the facility hitting a tungsten target and lose energy as they propagate.
Specifically, we choose two theoretical models predicting light, MeV-scale LLPs, namely, a light bino in the RPV-SUSY, and a weak-violating electrophilic ALP.
We consider production modes of the light bino including decays of $\pi^+ \to e^+ \tilde{\chi}^0_1$ and $\mu^+ \to e^+ \overset{(-)}{\nu} \tilde{\chi}^0_1$, where $\mu^+$ stems from the $\pi^+\to \mu^+ \nu_\mu$ decay at rest {and also decays at rest immediately}.
The light bino then travels and decays inside SHiNESS into an electron-positron pair in association with a neutrino.
Three benchmark scenarios, each with two non-vanishing RPV couplings mediating the production and decay of the light bino, respectively, are studied.
Included is also one benchmark where a single RPV coupling induces both the production and decay of the light bino.
Numerical results are then obtained and presented in the plane of either the production coupling vs.~the decay coupling, or the RPV couplings (assumed to be equal) versus the light bino's mass.
Our findings show that SHiNESS is predicted to be able to exclude large, new territories in the parameter space with just two years' data collection for the considered benchmark scenarios.

For the ALP, we focus on the $\pi^+\to e^+ \, \nu_e \, a$ decay for its production and on $a\to e^+ e^-$ for its signal decay.
We assume a single, weak-violating term in an effective Lagrangian with the ALP coupled to a pair of electrons, that, after integration by parts, leads to both a helicity-suppressed three-point-vertex term, and a helicity-unsuppressed four-point-vertex term.
The latter contribution, involving in the vertex an ALP, an electron, an electron neutrino, as well as a $W$-boson, results in large production rates of the ALP from the $\pi^+\to e^+ \nu_e a$ decay.
In the numerical results, we find that with solely two years' operation, SHiNESS can probe a unique, unexcluded part of the parameter space of $c_{ee}/\Lambda\sim \mathcal{O}(10^{-5})$ GeV$^{-1}$ for $m_a$ between 10 MeV and 120 MeV, complementary to other future experiments which would be sensitive to larger values of $c_{ee}/\Lambda$ in a similar range of $m_a$.

Before closing, we note that in addition to decays of the $\pi^+$ mesons,  decays of heavier pseudoscalar mesons such as $K^+$ can contribute to the LLP production.
In spite of the smaller production rates, kaons may allow to probe heavier LLPs for kinematic reasons.
We leave this possibility for future studies.

\section*{Acknowledgements} 
We thank Liangwen Chen, Martin Sch\"urmann, and Stefano Roberto Soleti, for useful discussions. Y.Z.~is supported by the National Natural Science Foundation of China under Grant No.~12475106 and the Fundamental Research Funds for the Central Universities under Grant No.~JZ2023HGTB0222.
W.L.~is supported by National Natural Science Foundation of China (Grant No.~12205153).

\bibliographystyle{JHEP}
\bibliography{refs}

\end{document}